\documentclass[prl,superscriptaddress,twocolumn,aps,longbibliography]{revtex4}
\usepackage{graphicx}% Include figure files
\usepackage{dcolumn}% Align table columns on decimal point
\usepackage{bm}% bold math
\usepackage[colorlinks]{hyperref}

\begin{document}

\preprint{APS/123-QED}

\title{Spatial-spectral mapping to prepare the frequency entangled qudits}

\author{Zi-Xiang Yang}
\affiliation{Hubei Key Laboratory of Optical Information and Pattern Recognition, Wuhan Institute of Technology, Wuhan 430205, China}

\author{Zi-Qi Zeng}
\affiliation{Hubei Key Laboratory of Optical Information and Pattern Recognition, Wuhan Institute of Technology, Wuhan 430205, China}

\author{Ying Tian}
\affiliation{Hubei Key Laboratory of Optical Information and Pattern Recognition, Wuhan Institute of Technology, Wuhan 430205, China}

\author{Shun Wang}
\affiliation{Hubei Key Laboratory of Optical Information and Pattern Recognition, Wuhan Institute of Technology, Wuhan 430205, China}

\author{\\ Ryosuke Shimizu } 
\affiliation{The University of Electro-Communications, 1-5-1 Chofugaoka, Chofu, Tokyo, Japan}

\author{Hao-Yu Wu}
\email{haoyuwu@wit.edu.cn}
\affiliation{Hubei Key Laboratory of Optical Information and Pattern Recognition, Wuhan Institute of Technology, Wuhan 430205, China}

\author{Shilong Liu}
\email{dr.shilongliu@gmail.com}
\affiliation{Department of Physics, University of Ottawa, 25 Templeton Street, K1N 6N5, Ottawa, ON, Canada}
\affiliation{femtoQ Lab, Engineering Physics Department, Polytechnique Montr\'{e}al, Montr\'{e}al, Qu\'{e}bec H3T 1JK, Canada}

\author{Rui-Bo Jin}
\email{jin@wit.edu.cn}
\affiliation{Hubei Key Laboratory of Optical Information and Pattern Recognition, Wuhan Institute of Technology, Wuhan 430205, China}
\affiliation{Guangdong Provincial Key Laboratory of Quantum Science and Engineering, Southern University of Science and Technology, Shenzhen 518055, China}

\date{\today}

\begin{abstract}
Entangled qudits, the high-dimensional entangled states, play an important role in the study of quantum information. 
 How to prepare entangled qudits in an efficient and easy-to-operate manner is still a challenge in quantum technology. 
Here, we demonstrate a method to engineer frequency entangled qudits in a spontaneous parametric downconversion process. 
The proposal employs an angle-dependent phase-matching condition in a nonlinear crystal, which forms a classical-quantum mapping between the spatial (pump) and spectral (biphotons) degrees of freedom. 
In particular, the pump profile is separated into several bins in the spatial domain, and thus shapes the down-converted biphotons into discrete frequency modes in the joint spectral space. 
Our approach provides a feasible and efficient method to prepare a high-dimensional frequency entangled state.
As an experimental demonstration, we generate a three-dimensional entangled state  by using a homemade variable slit mask.
\end{abstract}

\maketitle

\section{Introduction}
Quantum entangled states are serving as essential resources in quantum technologies, e.g., quantum computation \cite{Knill2001}, communications \cite{gisin2007quantum}, and measurements \cite{clerk2010introduction}.
The high-dimensional entangled states (or named as entangled qudits with a dimension number of $d$)   demonstrate significant progress 
%show many advances 
in the aforementioned quantum applications \cite{Erhard2020}.
In quantum communication, a high-dimensional quantum state could carry more information, thus increasing the channel capacities and also the noise resilience \cite{Chen2021,dada2011experimental}.
 In quantum computation, entangled  qudits not only have a larger state space to store and process information but also have the ability to do multiple control operations simultaneously \cite{Lu2019}. These features  are significant in the reduction of the circuit complexity and the acceleration of the algorithm \cite{ Reimer2018}.
In quantum measurement, a strong reduction in the number of operations can be achieved by using qudit systems satisfying a certain relation between their dimensionality and topology \cite{Giovannetti2004}. 
In addition,  entangled  qudits are crucial for studies of fundamental quantum mechanics, i.e., the non-locality in high-dimensional quantum systems \cite{dada2011experimental}.

%[Progress and challenges]
 The high-dimensional entangled state could be realized in various degrees of freedom of the photon, including time \cite{LagoRivera2021},
frequency \cite{Kues2017,Chen2021,Imany2018},
 hybrid time-frequency modes \cite{Reimer2018,Xiang2020},
 paths \cite{Lu2020}, and orbital angular momentum (OAM) \cite{liu2018coherent}.
Regarding the time-frequency entangled qudits, they are intrinsically suitable for long-distance transmission in optical fibers, waveguides, as well as free space. Therefore, it has attracted much attention in recent years \cite{Chen2021,  Lingaraju:19, PhysRevLett.123.123603, Bernhard2013,XieZhongShresthaEtAl2015,KuesReimerRoztockiEtAl2017, Jin2016QST,Jin2021APLphoton,Lixiaoying2020APL,Morrison2022APL,Yang2020} .

Generally speaking, the previous methods for the generation of the frequency-entangled qudits could be classified into four categories:
(1) using a linear pulse shaper, which may include two optical gratings and one spatial light modulator (SLM) \cite{Bernhard2013,Liu2022SM};
(2) utilizing a nonlinear cavity, such as a Fabry-Perot cavity or  ring-cavity  \cite{XieZhongShresthaEtAl2015,KuesReimerRoztockiEtAl2017};
(3) employing an interferometer, for example, the spectrally resolved Hong-Ou-Mandel(HOM) interferometer \cite{Jin2016QST}, quantum optical synthesis in a dual pump interferometer \cite{Jin2021APLphoton}, or a nonlinear interferometer in fibers \cite{Lixiaoying2020APL};
(4) developing quantum state engineering in a nonlinear material, such as a customized poling nonlinear crystal \cite{Morrison2022APL}.
All the above schemes require sophisticated modulation devices or specially designed crystals, which may lead to a large loss or high cost during the preparation. 
Therefore, it's meaningful to explore one simple and efficient method to generate a high-dimensional frequency entangled state.

%
%=============================================
\begin{figure*}[!ht]
\centering
\includegraphics[width=0.75\textwidth]{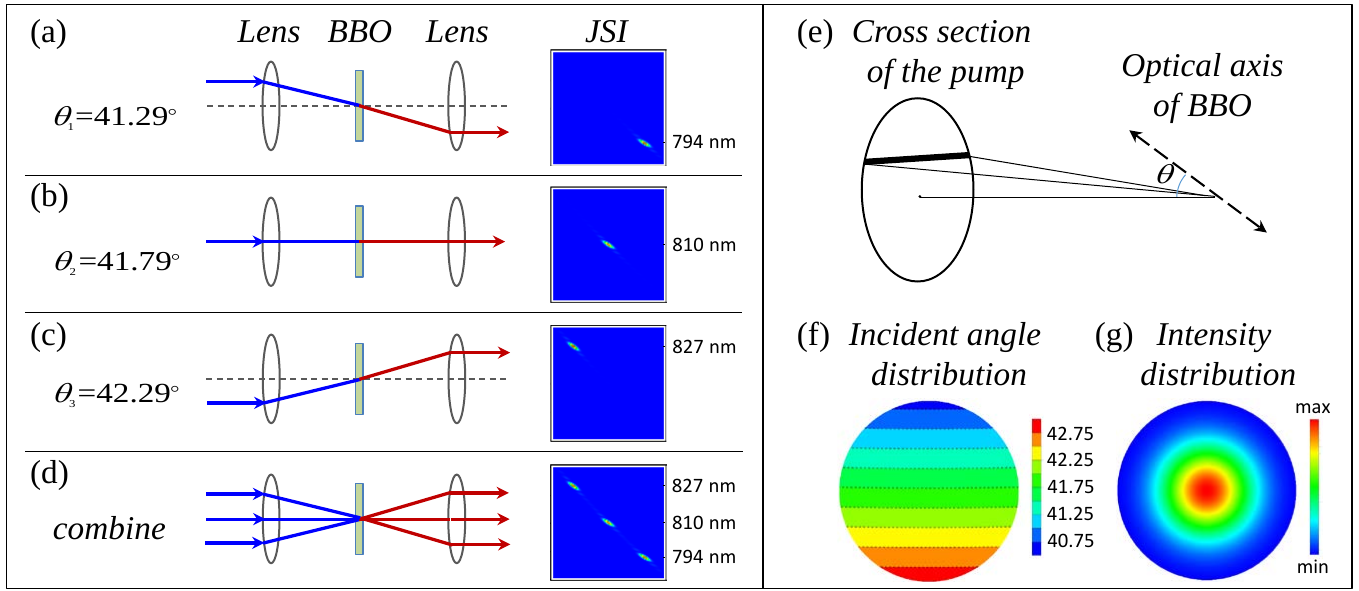}
\caption{The principle of spatial-spectral mapping. (a-c) The different incident angles of the pump beam correspond to discrete spectral modes. (d) The concept of combining three beams to prepare a three-dimensional entangled state. (e) The configuration of the incident angle $\theta$: the angle between the pump laser and the optical axis of the crystal. (f) The  incident angle distribution on the cross section of the pump beam. (g) The intensity distribution on the cross section of the pump beam.}
\label{concept}
\end{figure*}
%=============================================
%
%
%=============================================
\begin{figure*}[!ht]
\centering
\includegraphics[width=0.75\textwidth]{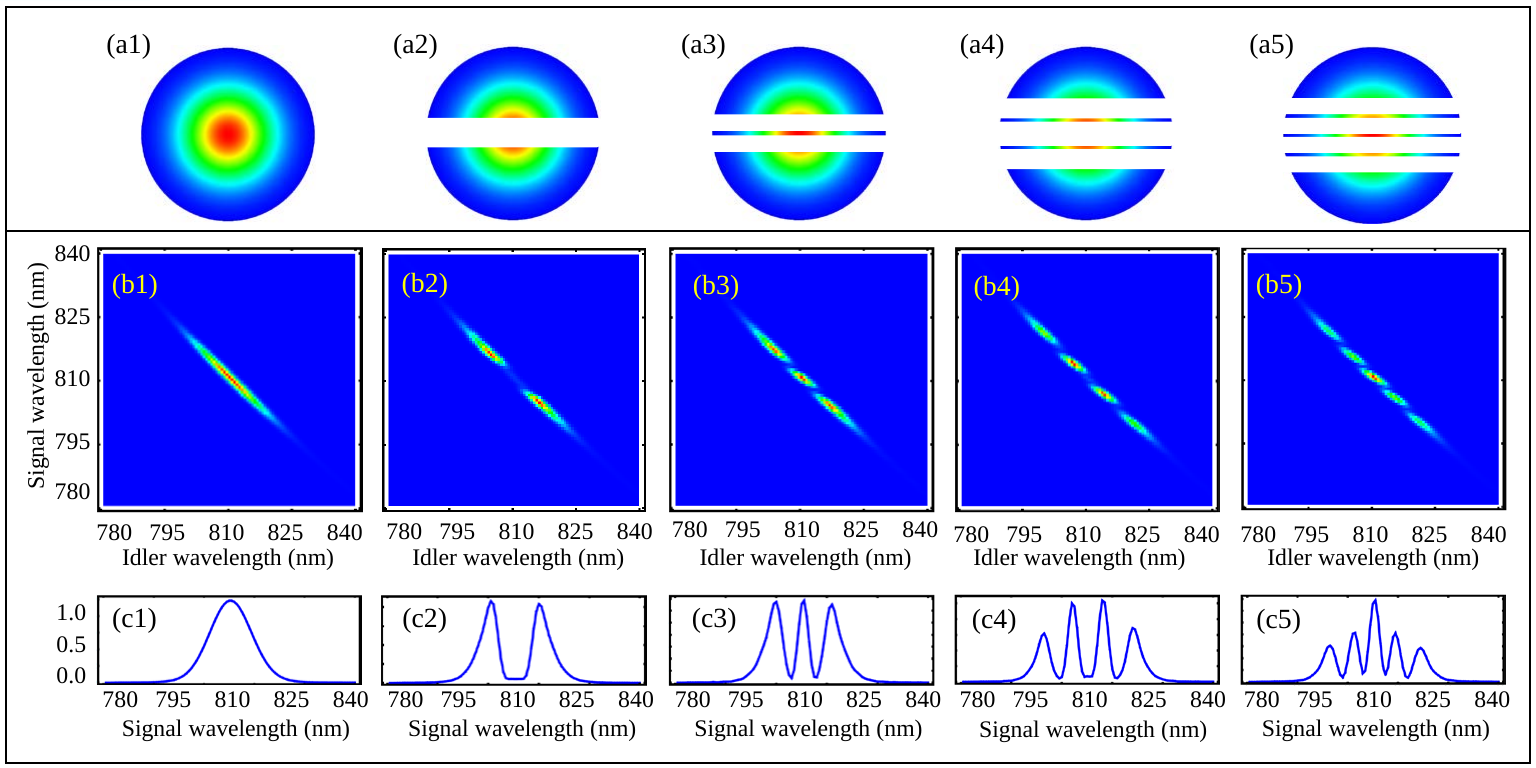}
\caption{Simulation of the spatial-spectral mapping to prepare high-dimensional entangled states.
(a1-a5): The cross section of the pump beam, spatially filtered by different masks.
(b1-b5): The calculated JSIs under the corresponding spatial masks.
(c1-c5): The spectra of the signal photon are obtained by projecting the JSI onto the vertical  axis. See supplemental document for more simulations.}
\label{simulation}
\end{figure*}
%=============================================
%

Here, we propose and experimentally demonstrate a feasible method to generate frequency-entangled qudits using a $\beta$-barium borate (BBO) crystal. This is possible by employing two improvements in our regime. On the one hand, we employ an angle-dependent phase-matching (also called birefringent phase-matching) condition  of the BBO crystal, to form a classical-quantum mapping between the spatial (pump) and spectral
(photon pairs) degrees of freedom. On the other hand, we build a homemade spatial mask, which could make a spatial modulation on the pump profile. For example, it could separate the pump into several spatial bins and thus on-demand control the entangled spectral modes in the joint spectral space. The proposed regime is of great feasibility and provides one efficient way to prepare the high-dimensional entangled qubit in the frequency domain.

%
%=============================================
\begin{figure*}[!tb]
\centering
\includegraphics[width=0.75\textwidth]{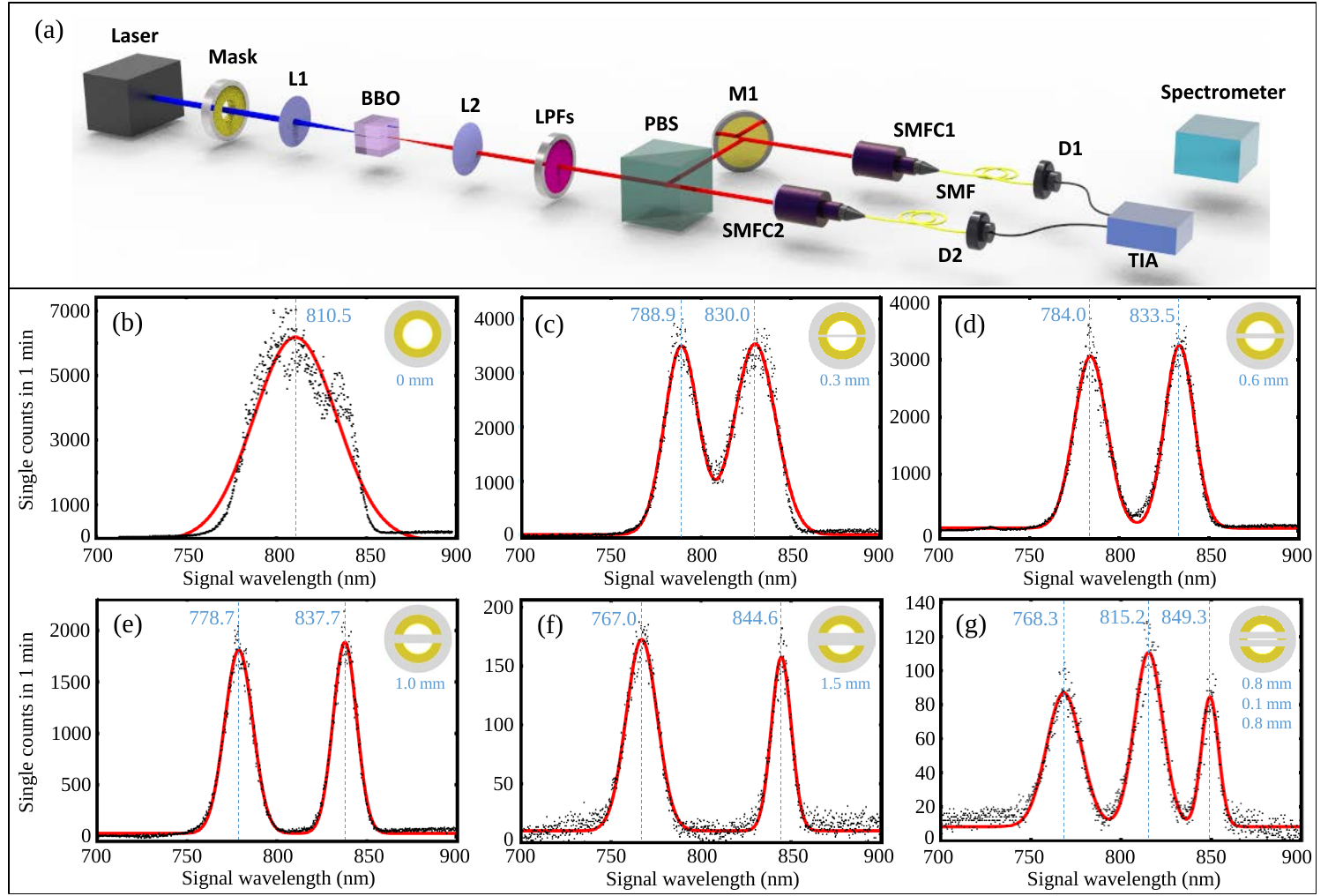}
\caption{
(a) The experimental setup and results for preparation of the frequency entangled qudits.  L1(2): lens, LPFs: long-pass filters,
PBS: polarizing beam splitter,
M1: mirror,
D1(2): detector, TIA: time interval analyzer,  SMFC: single-mode-fiber coupler.
(b-g) The measured spectra of the signal photon using different masks, which are shown in the upper-right corner of each figure.  }
\label{experiment}
\end{figure*}
%=============================================
%

\section{Theory and simulation}
The biphoton state  generated in a spontaneous parametric down conversion (SPDC) process can be expressed as\cite{Mosley2008PRL}:
\begin{equation}\label{eq:1}
\left| \psi  \right\rangle  = \int_0^\infty  {\int_0^\infty  {d{\omega _s}d{\omega _i}f({\omega _s},{\omega _i},\theta )} } {\hat a^\dag }_s({\omega _s}){\hat a^\dag }_i({\omega _i})\left| 0 \right\rangle,
\end{equation}
where the subscripts $s$ and $i$ represent the signal and idler photon, respectively; $\omega $ is the angular frequency; ${\hat a^\dag }$ is the creation operator; $f({\omega _s},{\omega _i},\theta )$ is the joint spectral amplitude (JSA) of the biphoton, where $\theta$ is the incident angle of the pump, i.e., the angle between the pump laser and the optical axis of the crystal (See Fig.\,\ref{concept} (e)).
$f({\omega _s},{\omega _i},\theta )$ is the product of the pump-envelope  function $\alpha ({\omega _s, \omega _i})$ and the phase-matching function $\phi ({\omega _s},{\omega _i, \theta})$ \cite{Mosley2008PRL}, i.e.,
\begin{equation}\label{eq:2}
 f({\omega _s},{\omega _i},\theta ) = \alpha ({\omega _s},{\omega _i}) \times \phi ({\omega _s},{\omega _i},\theta ).
\end{equation}
The widely used pump-envelope function  is a  Gaussian function, i.e.,
\begin{equation}\label{eq:3}
\alpha ({\omega _s},{\omega _i}) = \exp \left[ - \frac{1}{2}{\left(\frac{{{\omega _s} + {\omega _i} - {\omega _p}}}{{{\sigma _p}}}\right)^2}\right],
\end{equation}
where $\omega _p$ and $\sigma _p$ are the central frequency and bandwidth of the pump, respectively. 
The phase-matching function in a nonlinear crystal can be expressed as \cite{Mosley2008PRL}
\begin{equation}\label{eq:4}
\phi ({\omega _s},{\omega _i},\theta ) = {\rm{sinc}} \left(\frac{{\Delta kL}}{2}\right)\exp \left(\frac{{i\Delta kL}}{2}\right),
\end{equation}
where $L$ is the crystal length; $\Delta k = {k_s} + {k_i} - {k_p}$ is the  wave vector  difference between the signal, idler, and pump, respectively.
The wave vector is a function of the incident angle $\theta$.

The principle for generating qudits is shown in Fig.\,\ref{concept}.
Thanks to the angle phase-matching (also called birefringent phase-matching) condition, the different incident angle $\theta_j$  corresponds to the different discrete spectral mode in the joint spectral intensity (JSI) $|f({\omega _s},{\omega _i},\theta_j ){|^2}$.
For example, Fig. \ref{concept} (a-c) shows the cases of $\theta_j=41.29^{\circ}$, $41.79^{\circ}$, and  $42.29^{\circ}$, respectively. 
The corresponding central wavelengths are  794 nm (827 nm), 810 nm (810 nm), and 827 nm (794 nm) for the signal (idler) photon, respectively.
If we shape the pump profile to $d$ spatial rays, the output state will become a separate but superposed $d$-dimensional entangled state:
\begin{equation}% \label{}
  \left| \psi  \right\rangle=\sum_{j=1}^{d}c_j \left| \omega_s, \omega_i  \right\rangle_j=\sum_{j=1}^{d}c_j \left| j\right\rangle_{\omega_s}  \left| j\right\rangle  _{\omega_i}, 
\end{equation}
Here, $c_j$ is the coefficient that depends on both the spatial pump beam and JSA (JSI). In this case, the JSI could be expressed as:
\begin{equation}\label{eq:5}
JSI = \left|  \sum_{j=1}^{d}  f({\omega _s},{\omega _i},\theta_j ) \right|^2.
\end{equation}
As a demonstration, Fig.\,\ref{concept}(d) shows three-dimensional entangled qudits.

Under the paraxial approximation, the incident angle of the pump beam is only determined by the vertical axis. In other words,  all the positions on the same horizontal line have the same incident angle, as shown in Fig.\,\ref{concept}(e).
This phenomenon can be intuitively understood in the following way:
since the focusing range (e.g., 50 mm in our scheme) is much larger than the focused beam radius (less than 1 mm) on the BBO crystal, the contribution from the horizontal angle can be neglected.
Fig.\,\ref{concept}(f, g) shows the incident angle and intensity distribution on the cross section of the pump beam.

Based on the incident angle distribution in Fig.\,\ref{concept}(f),   we can shape the pump beam using homemade masks. 
Fig.\,\ref{simulation}(a1-a5) and (b1-b5) show the simulations, where the different masks correspond to the unique JSIs.
When no slit impinges on the pump beam, a tilt and long JSI can be generated, as shown  in Fig.\,\ref{simulation}(a1, b1).
When the center position of the pump beam is blocked using a mask in Fig.\,\ref{simulation}(a2), the center position in the JSI is lost and a two-mode frequency entangled state is generated, as shown in Fig.\,\ref{simulation}(b2).
When the two sections between the center are blocked using the mask in Fig.\,\ref{concept}(a3), two related sections in the JSI are also lost, resulting in a  three-dimensional qudit in Fig.\,\ref{concept}(b3).
Following a similar way, four- and five-dimensional qudits in Fig.\,\ref{simulation}(b4, b5) could also be generated using three- and four-line masks in Fig.\,\ref{simulation}(a4, a5).

By projecting the two-dimensional distribution diagram of JSI onto the vertical and horizontal   axes, the spectral distribution of the signal and idler photon can be obtained.
Fig.\,\ref{simulation}(c1-c5) shows the corresponding distribution of the signal photon, which has a distribution of 1, 2, 3, 4, and 5 peaks. 
The full-width at half-maximum (FWHM)  of the peak in Fig.\,\ref{simulation}(c1) is  19.7 nm.
Note by adjusting the width and the center position of the mask, different JSIs and spectra can be prepared. See the supplemental document for more simulations.
Next, we verify this scheme in an experiment.

\section{Experiment and Results}

%\section{Experiment}
The experimental setup for generating   frequency entangled qudits is shown in Fig.\,\ref{experiment} (a).
The laser used in this experiment is a single-transverse-mode and multi-longitudinal-mode laser diode (LD), which has a central wavelength of 405 nm and a bandwidth of 0.53 nm \cite{Cai2022}. The laser beam (with a diameter of around 2 mm) is spatially filtered by a mask and then focused using a lens (L1, f=50 mm). The 5-mm-long BBO crystal is  designed for a  type-II  phase-matched (e$\rightarrow$o+e)  SPDC at 810 nm (cut angle: $\theta  = 41.9^\circ$  and  $ \varphi  = 0^\circ $).
% .
The downconverted biphotons are collimated by the second lens (L2, f=50 mm) and then filtered by a set of long-pass filters (LPFs). Here, we define the  reflected (transmitted) photon from the PBS as the signal (idler).
After being separated by a polarization beam splitter (PBS), the biphotons are coupled into the single-mode fibers (SMFs),  which are connected to two single-photon detectors D1 and D2 (SPCM-AQRH-10-FC from Excelitas) and a time interval analyzer (TIA, Picoharp300 from Pico Quanta Co.). When the pump power is set to 40 mW, we obtain single counts of 762 kHz for the signal and 325 kHz for the idler. Here, the gap in single counts is due to the different coupling efficiencies of each channel. The measured coincidence count is 4.1 kHz.

After the photon counting test, the biphotons' spectra are measured by a single-photon level spectrometer (SP2300, Princeton Instrument Co.).
Figure\,\ref{experiment}(b) is the spectrum of the signal photon using a homemade  mask with no block line inside. The FWHM of this spectrum is 54.2 nm.
This mask is made of photocurable resin using a 3D printer. 
By inserting a mask with one block line, we obtain the spectrum of the two-dimensional qudits, as shown in Fig.\,\ref{experiment}(c, d, e, f). 
To investigate the orthogonality of the frequency qudits, we set  the width of the center block line as 0.3 mm, 0.6 mm, 1.0 mm, and 1.5 mm in Fig.\,\ref{experiment}(c, d, e, f),  respectively.
It can be noticed that  the two peaks in Fig.\,\ref{experiment}(c, d) are partially overlapped, while the two peaks in Fig.\,\ref{experiment}(e, f)  are completely separated, indicating that the two frequency components are  orthogonal. The spacing between the two peak centers is 41.1 nm, 49.5 nm, 59.0 nm, and 77.6 nm, respectively. 
Figure\,\ref{experiment}(g) depicts the spectrum of the three-dimensional entangled qudits using a mask with two block lines inside. The width of each block line is 0.8 mm and the spacing between the two lines is 0.1 mm. The three frequency components in Fig\,\ref{experiment}(g)  are also well separated.

\section{Conclusion}
In conclusion, the present scheme may provide a useful platform to engineer the more complex frequency entangled state. For example, by employing a flattop spectrum of the pump, the regime could generate a high-dimensional maximum entangled state \cite{liu2020increasing}. 
By selecting one spatial mask with a high resolution, i.e., an SLM, the dimensional space of the entangled state could be further improved. Also, a high-dimensional Bell state may be possible to perform with the help of the pump modulation technique \cite{liu2018coherent}.
 In addition, the reported spatial-spectral mapping scheme is beneficial to study the spatial-temporal entangled states \cite{Brecht2015PRX}.

\section*{Acknowledgments}
This work was supported by the National Natural Science Foundations of China (Grant Numbers 91836102, 12074299,  11704290, and 11904112); Guangdong Provincial Key Laboratory (Grant No. GKLQSE202102) and Natural Science Foundation of Hubei Province (2022CFA039).

%\bibliography{MappingarXiv}

\clearpage
\newpage
\onecolumngrid

\renewcommand\thefigure{A\arabic{figure}}
\setcounter{figure}{0}

\setcounter{equation}{0}
\renewcommand\theequation{A\arabic{equation}}
\setcounter{page}{1}

%\section{Supplementary Information}

%\section*{Appendix}

\section*{Spatial-spectral mapping to prepare the frequency entangled qudits: supplemental document }
%\setboolean{displaycopyright}{false} %copyright statement should not display in the  supplementary document

\renewcommand\thefigure{S\arabic{figure}}
\setcounter{figure}{0}

\setcounter{equation}{0}
\renewcommand\theequation{S\arabic{equation}}

\begin{figure}[ht]
\centering
\includegraphics[width= 0.95\textwidth]{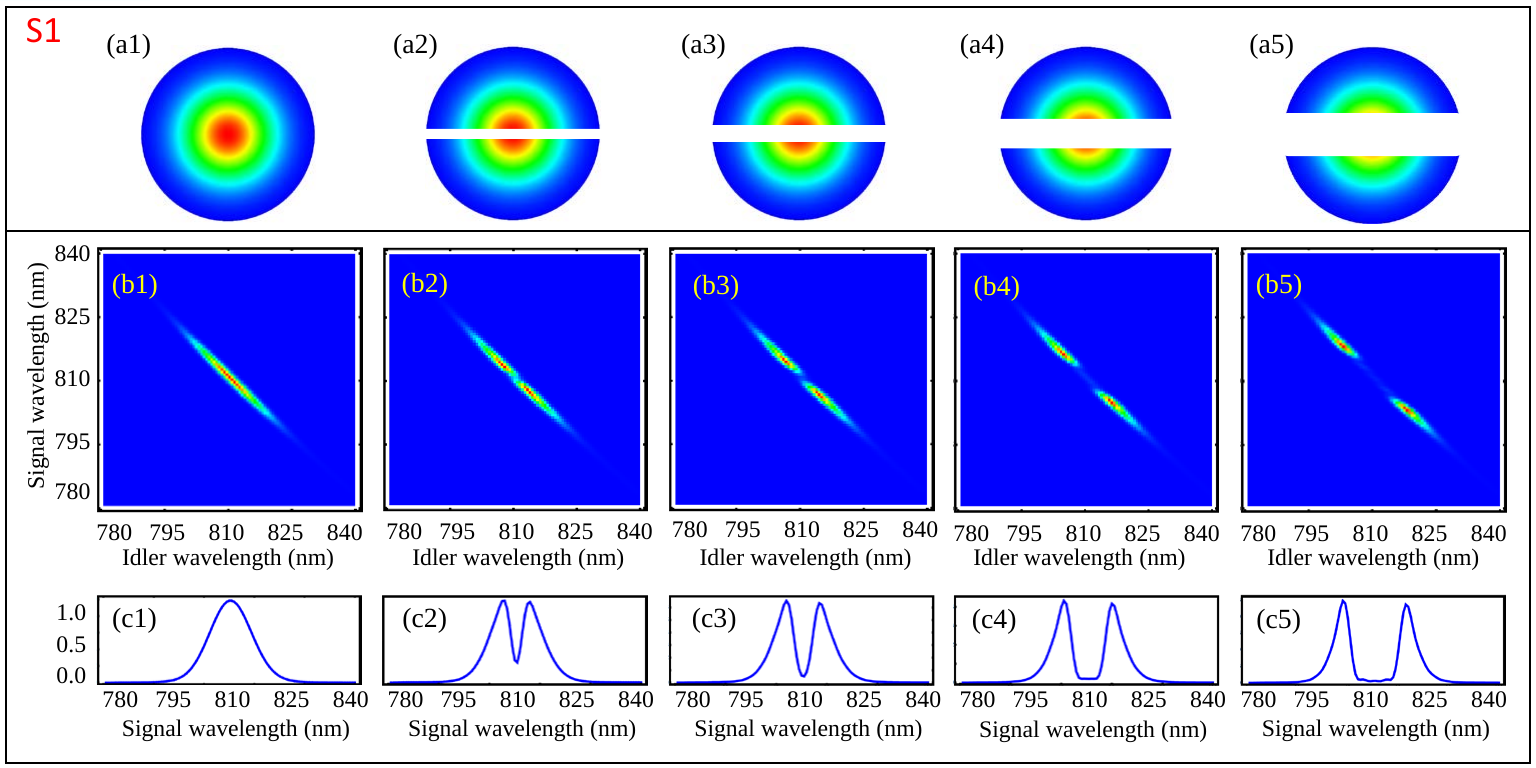}
\caption{Simulation of the spatial-spectral mapping to prepare high-dimensional entangled states.
(a1-a5): The cross section of the pump beam, spatially filtered by a single slit with different widths.
 The  beam is divided into 41 bins, and the  slit widths in (a1-a5) are 0, 3, 5, 9, 13 bins respectively.
(b1-b5): The calculated JSIs under the corresponding spatial masks.
(c1-c5): The spectra of the signal photon are obtained by projecting the JSI onto the vertical  axis.}
\label{Fig-S1}
\end{figure}

%\section*{S4: Comparison of the simulation and experimental results}
In this supplemental document, we present the simulations of joint spectral intensity (JSI) and the spectra of biphotons by  adjusting the width and the center position of the mask.
Figure \,\ref{Fig-S1} shows the case of changing the width of a single slit.
%
%=============================================
%=============================================

\begin{figure}[ht]
\centering
\includegraphics[width= 0.95\textwidth]{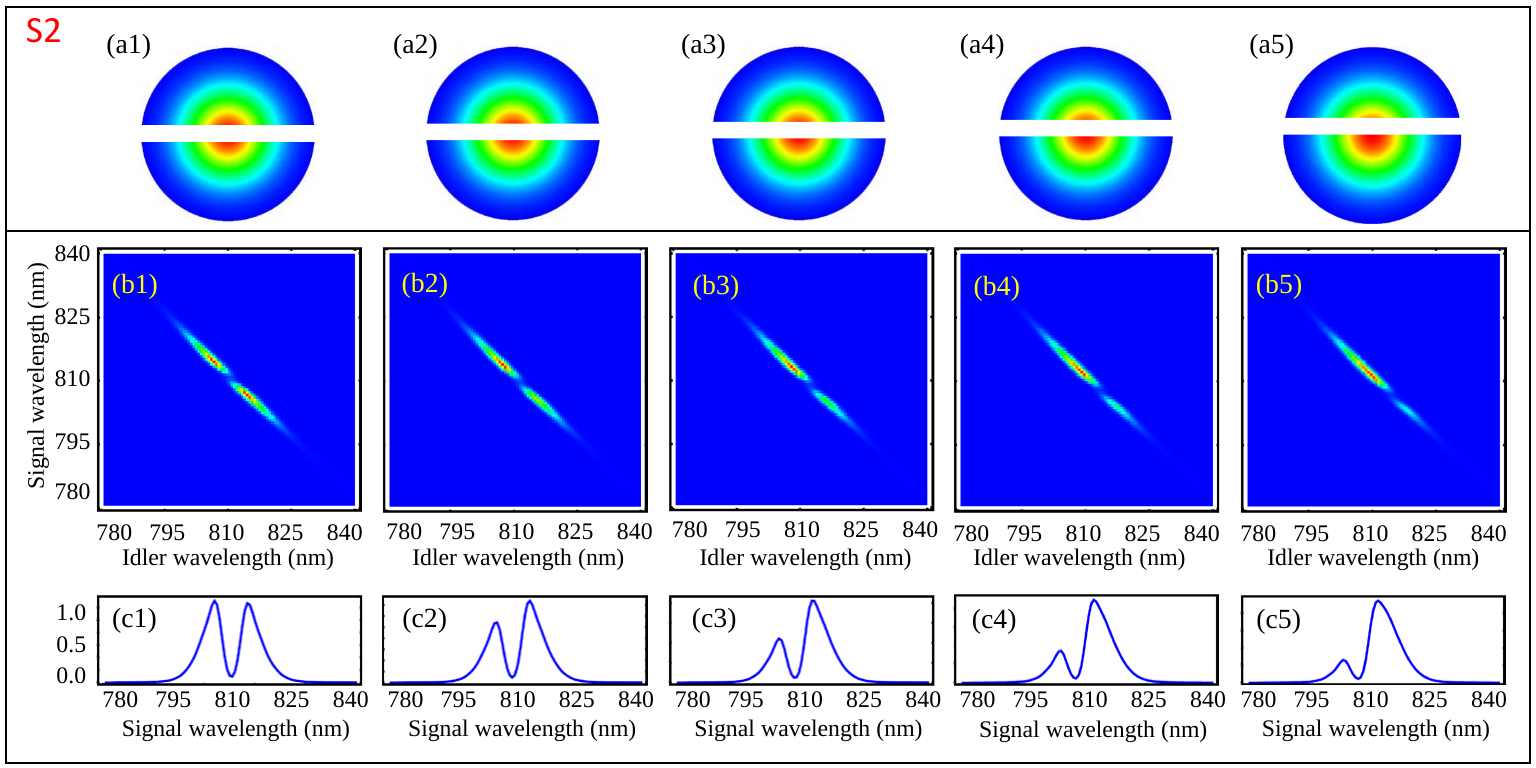}
\caption{Simulation of the spatial-spectral mapping to prepare high-dimensional entangled states.
(a1-a5): The cross section of the pump beam, spatially filtered by a single slit at different positions.
 The  beam is divided into 41 bins;  the width of the slit is 5 bins;  and the center positions of the slit   in (a1-a5) are 21, 22,  23,  24, and 25, respectively.
(b1-b5): The calculated JSIs under the corresponding spatial masks.
(c1-c5): The spectra of the signal photon are obtained by projecting the JSI onto the vertical  axis.}
\label{Fig-S2}
\end{figure}

Figure \,\ref{Fig-S2} shows the case of changing the position of a single slit (moving up).

%
%=============================================
%=============================================

\begin{figure}[ht]
\centering
\includegraphics[width= 0.95\textwidth]{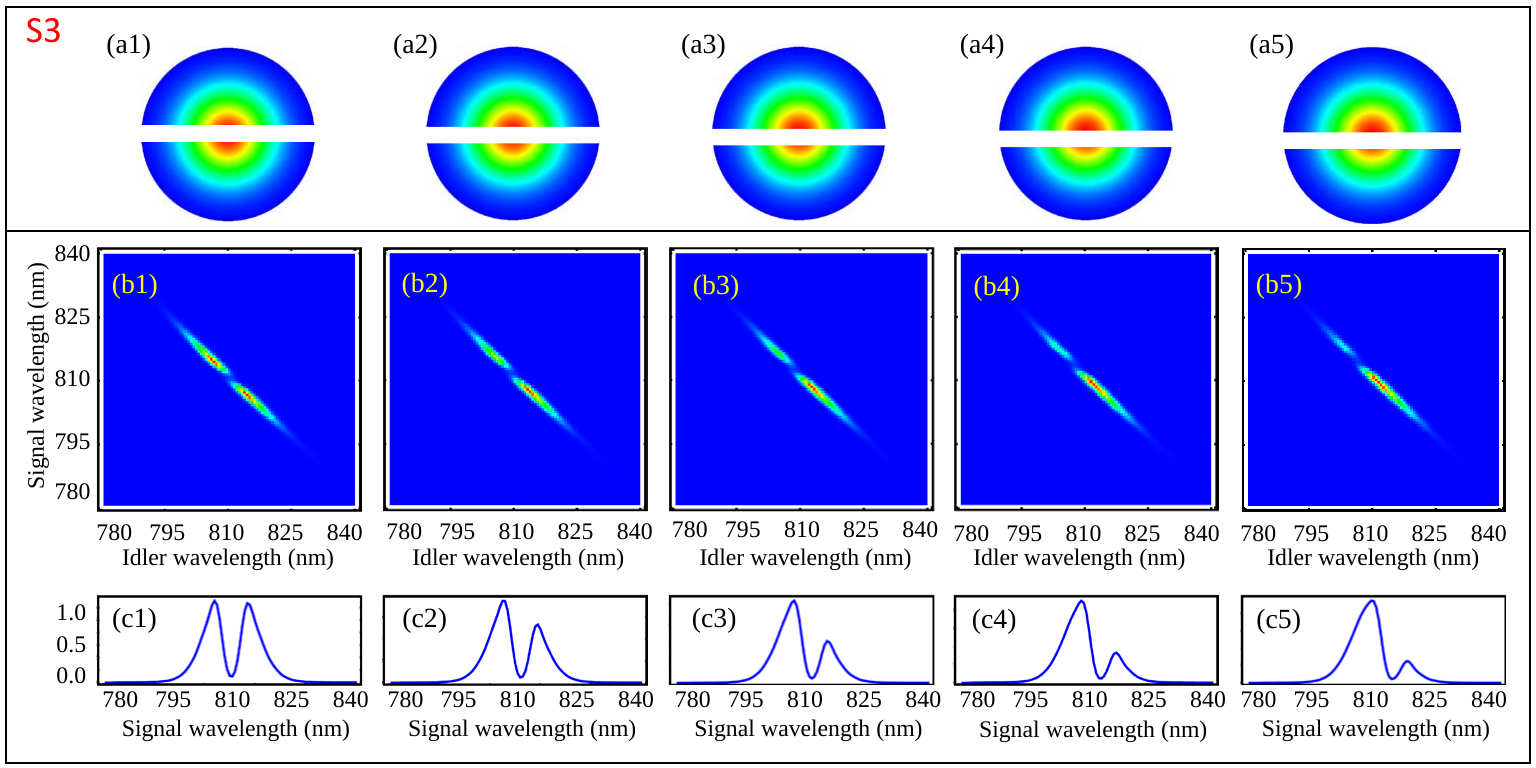}
\caption{Simulation of the spatial-spectral mapping to prepare high-dimensional entangled states.
(a1-a5): The cross section of the pump beam, spatially filtered by a single slit at different positions.
The  beam is divided into 41 bins;   the width of the slit is 5 bins; and  the center positions of the slit   in (a1-a5) are 21,  20,  19,  18, and 17, respectively.
(b1-b5): The calculated JSIs under the corresponding spatial masks.
(c1-c5): The spectra of the signal photon are obtained by projecting the JSI onto the vertical  axis.}
\label{Fig-S3}
\end{figure}

Figure \,\ref{Fig-S3}  shows the case of changing the position of a single slit (moving down).
%
%=============================================
%=============================================

\begin{figure}[ht]
\centering
\includegraphics[width= 0.95\textwidth]{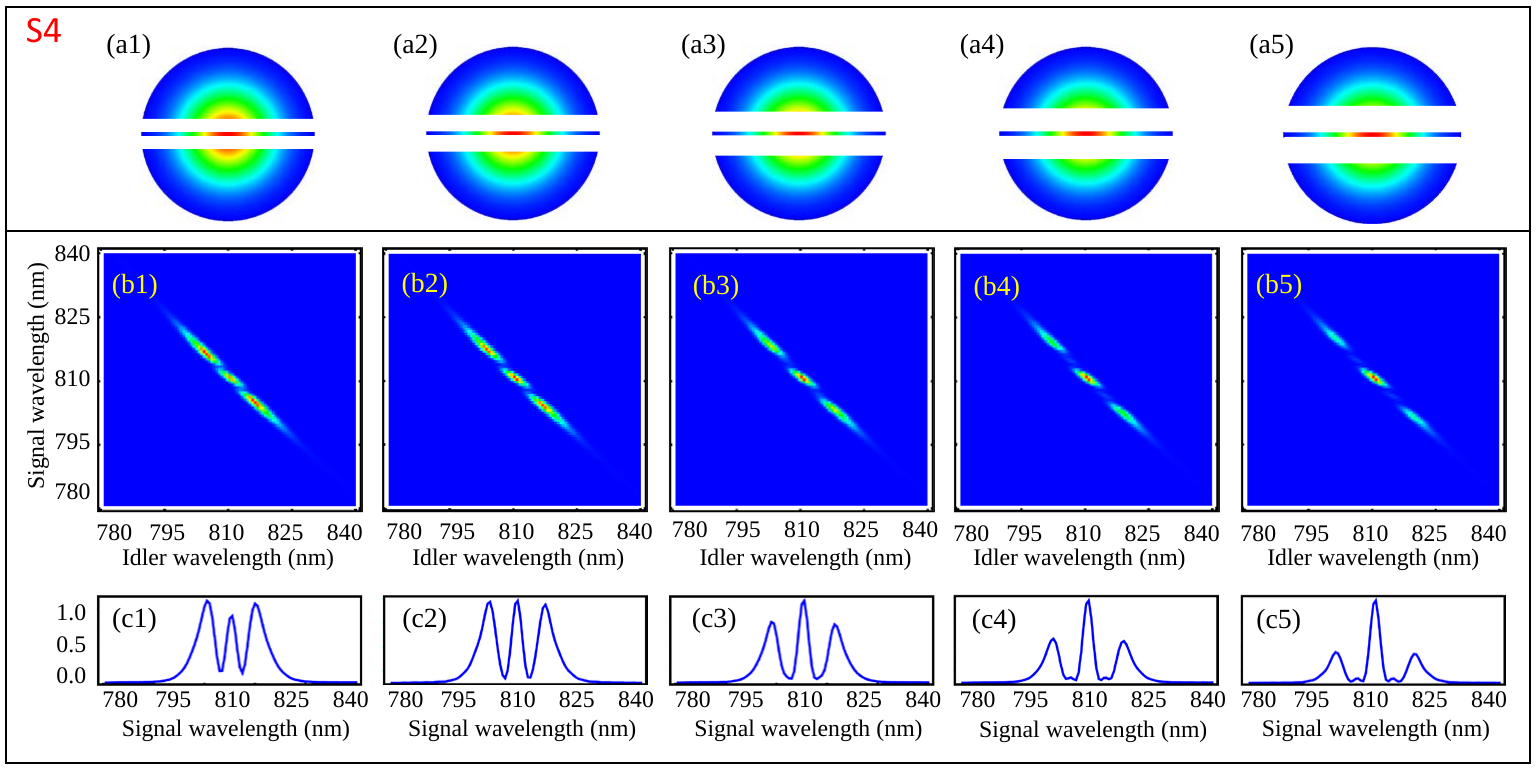}
\caption{Simulation of the spatial-spectral mapping to prepare high-dimensional entangled states.
(a1-a5): The cross section of the pump beam, spatially filtered by two slits with different widths and positions.
The  beam is divided into 41 bins;  the slit-beam-slit widths are 4-1-4, 5-1-5, 6-1-6, 7-1-7, 8-1-8, respectively.
(b1-b5): The calculated JSIs under the corresponding spatial masks.
(c1-c5): The spectra of the signal photon are obtained by projecting the JSI onto the vertical  axis.}
\label{Fig-S4}
\end{figure}

Figure \,\ref{Fig-S4} shows the case of changing the width and position of two slits.
%
%=============================================
%=============================================

\begin{figure}[ht!]
\centering
\includegraphics[width= 0.95\textwidth]{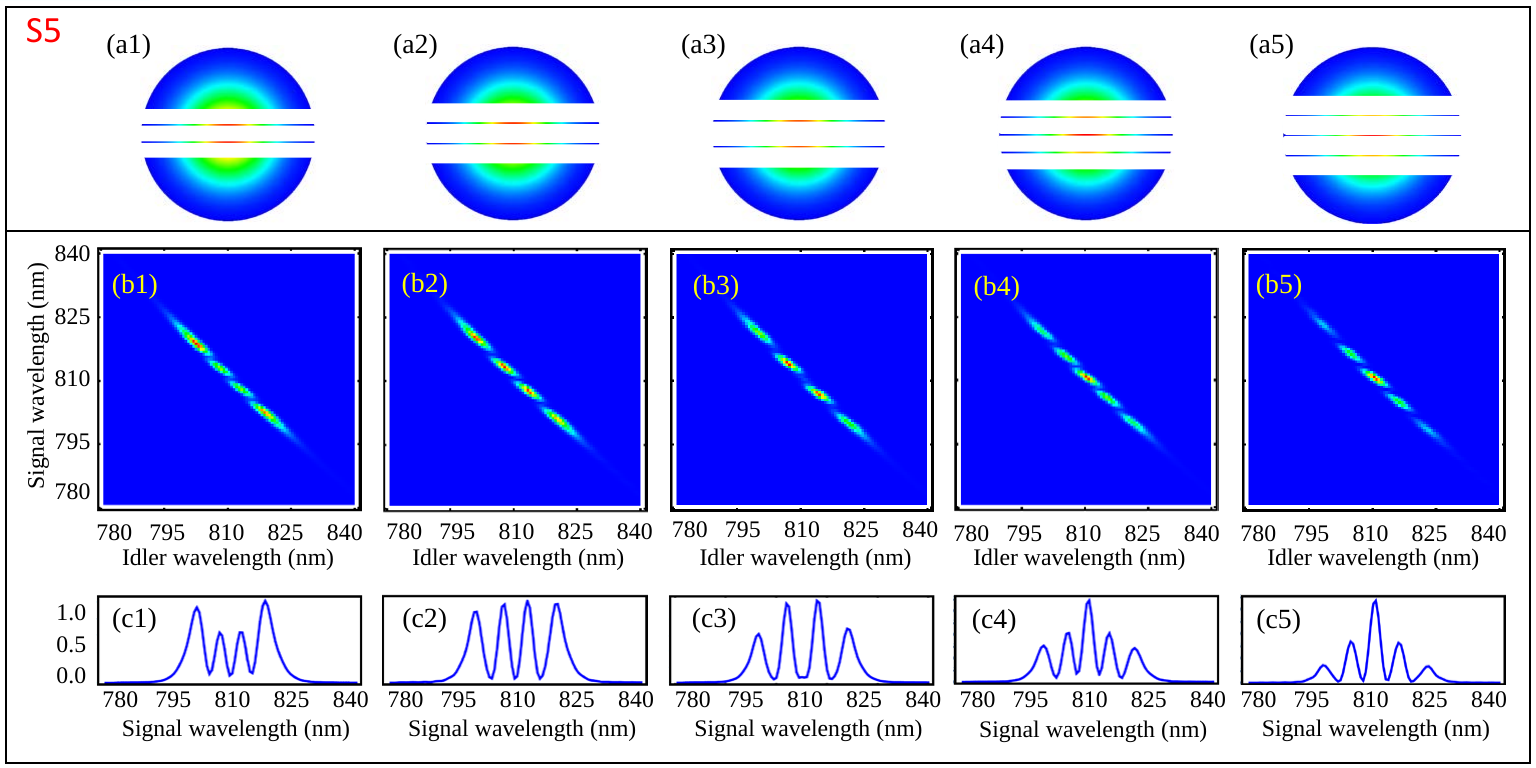}
\caption{Simulation of the spatial-spectral mapping to prepare high-dimensional entangled states.
(a1-a5): The cross section of the pump beam, spatially filtered by three or four slits  with different widths and positions.
  In (a1, a2, a3), the  beam is divided into 101 bins;  the slit-beam-slit-beam-slit widths are 11-2-11-2-11, 13-2-13-2-13, 15-2-15-2-15, respectively.
  In (a4), the  beam is divided into 41 bins;   the slit-beam-slit-beam-slit-beam-slit width is 4-1-4-1-4-1-4.
  In (a5),  the  beam is divided into 101 bins;  the slit-beam-slit-beam-slit-beam-slit width is 12-2-12-2-12-2-12.
(b1-b5): The calculated JSIs under the corresponding spatial masks.
(c1-c5): The spectra of the signal photon are obtained by projecting the JSI onto the vertical axis.}
\label{Fig-S5}
\end{figure}

Figure \,\ref{Fig-S5} shows the case of changing the width and position of three or four slits.
%
 %
%=============================================
%=============================================

\begin{figure}[ht]
\centering
\includegraphics[width= 0.95\textwidth]{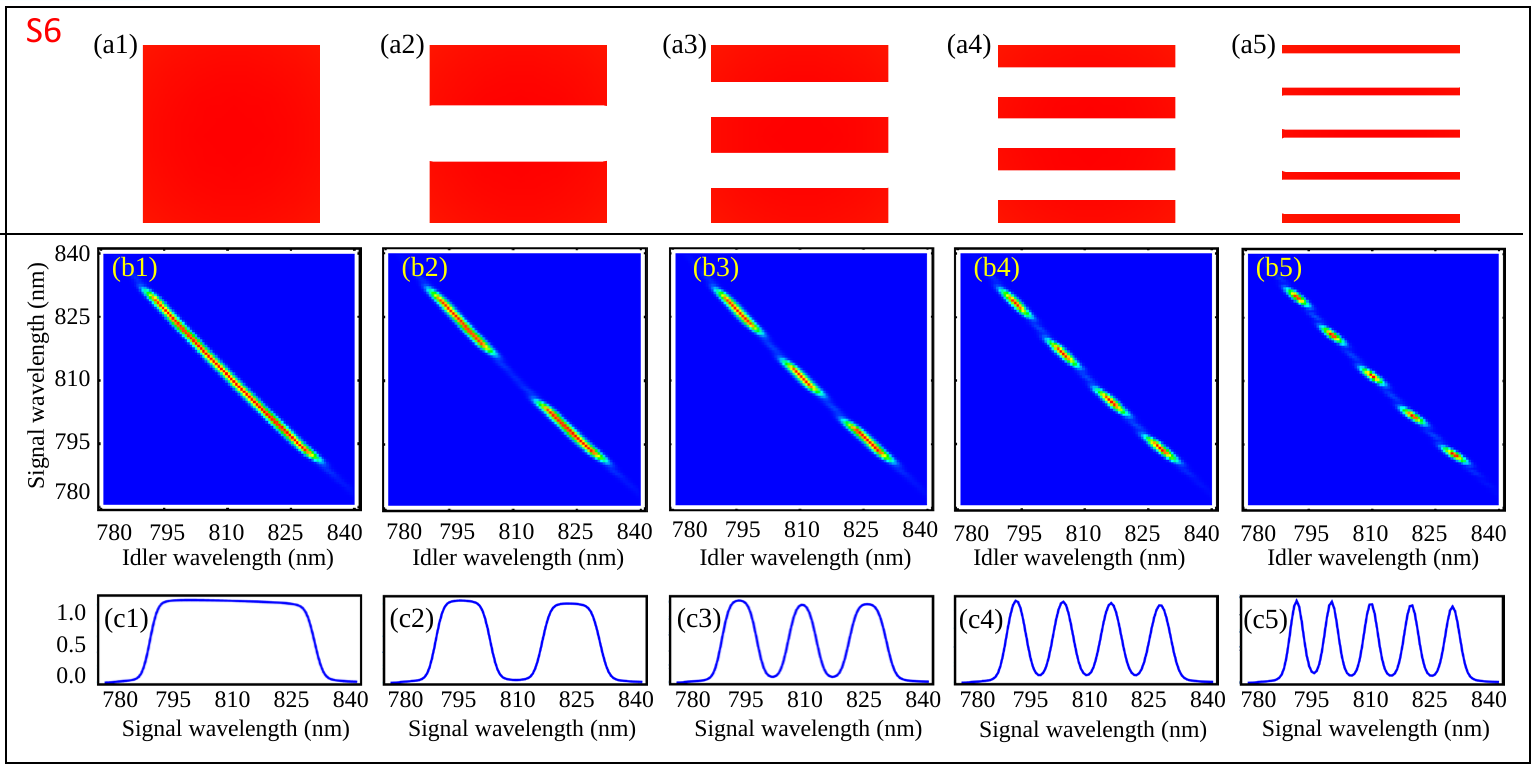}
\caption{
Simulation of the spatial-spectral mapping to prepare high-dimensional entangled states.
(a1-a5): The cross section of the pump beam with a flat-top distribution, spatially filtered by different slits.
The  beam is divided into 41 bins;  the slit and beam widths are 41,  14-13-14, 9-7-9-7-9,  5-7-5-7-5-7-5, 2-8-2-8-2-8-2-7-2, respectively.
(b1-b5): The calculated JSIs under the corresponding spatial masks.
(c1-c5): The spectra of the signal photon are obtained by projecting the JSI onto the vertical  axis.}
\label{Fig-S6}
\end{figure}

Figure \,\ref{Fig-S6} shows the case of utilizing a pump beam  with a flat-top distribution  rather than  a Gaussian distribution.
%
 %
%=============================================
%=============================================

%
\end{document}